\documentclass[a4paper,prd,twocolumn,aps,preprintnumbers,showpacs]{revtex4}
\usepackage{amssymb}
\usepackage{epsfig}
\usepackage{graphicx}% Include figure files
\usepackage{dcolumn}% Align table columns on decimal point
\usepackage{bm}% bold math
\usepackage{hyperref}
 
\def \be  {\begin{equation}}
\def \ee  {\end{equation}}
\def \ba  {\begin{eqnarray}}
\def \ea  {\end{eqnarray}}
\def \baa {\begin{eqnarray*}}
\def \eaa {\end{eqnarray*}}
\def \bb  {\begin{thebibliography}}
\def \eb  {\end{thebibliography}}

\def \nn {\nonumber}

\def \lab #1 {\label{#1}}

\def \fracs #1#2 {\mbox{\small $\frac{#1}{#2}$}}

\def \bin #1#2 {{\left({#1}\atop{#2}\right)}}
\def\lapproxeq{{\ \lower 0.6ex \hbox{$\buildrel<\over\sim$}\ }}
\def\gapproxeq{{\ \lower 0.6ex \hbox{$\buildrel>\over\sim$}\ }}

\def\hepph  #1 {{hep-ph/#1 }}

\begin{document}
%
%\preprint{XXXX}

\title{Jet production in (un)polarized $pp$ collisions: 
dependence on jet algorithm}
\author{Asmita Mukherjee$^{a,b}$ and Werner Vogelsang$^b$}
\affiliation{$^a$Department of Physics, Indian Institute of Technology Bombay, 
Powai, Mumbai 400076, India\\$^b$Institute for Theoretical Physics, 
T\"ubingen University, Auf der Morgenstelle 14, 72076 T\"ubingen, Germany}
%%%%%%%%%%%%%%%%
\begin{abstract}
We investigate single-inclusive high-$p_T$ jet production in longitudinally 
polarized $pp$ collisions at RHIC, with particular focus on the algorithm adopted to 
define the jets. Following and extending earlier work in the literature, we treat the 
jets in the approximation that they are rather narrow, in which case analytical results 
for the corresponding next-to-leading order partonic cross sections can be obtained. This approximation is 
demonstrated to be very accurate for practically all relevant situations, 
even at Tevatron and LHC energies. We confront results for cross sections and 
spin-asymmetries based on using cone- and $k_t$-type jet algorithms. We find
that jet cross sections at RHIC can differ significantly depending on the algorithm chosen, 
but that the spin asymmetries are rather robust. Our results are also useful for
matching threshold-resummed calculations of jet cross sections to fixed-order ones.
\end{abstract}

\date{\today}
\maketitle

%%%%%%%%%%%%%%%%%%%%%%
\section{Introduction}
%%%%%%%%%%%%%%%%%%%%%%
Jets are copiously produced at high-energy hadron colliders. Among other
things, they play important roles as precision probes of QCD and nucleon 
structure. At the Relativistic Heavy Ion Collider (RHIC), jets are by now  
a well-proven tool for investigating the spin structure of the nucleon,
through double-helicity asymmetries measured in the reaction
$pp\to {\mathrm{jet}}\,X$. The corresponding measurements~\cite{star} 
have in particular provided exciting information on the proton's polarized 
gluon distribution, $\Delta g$. 

There is no unique way of defining a jet. 
As a result, different jet algorithms exist and are being used in experiment. 
The jet definitions and algorithms can be broadly divided into two classes~\cite{soyez}: 
(i) successive recombination algorithms~\cite{catani,catani1,CA,css}, 
and (ii) cone algorithms~\cite{cone}. For the former, 
one first defines a distance between a pair of produced objects (initially, 
two particles), as well as a ``beam distance'' of each object to the collider beam axis. 
For each object, the smallest of these distances is determined. If it is a beam distance, 
the object is called a jet and removed from the list of objects in the event; 
otherwise the two objects are combined into a single object. This procedure is
repeated until no further recombinations take place. Prominent examples of 
successive recombination algorithms are the so-called $k_t$~\cite{catani,catani1}, 
Cambridge/Aachen~\cite{CA}, and anti-$k_t$~\cite{css}
algorithms, which differ in how the distances are defined.
We will collectively refer to such algorithms as ``$k_t$-type'' algorithms. 

Cone algorithms also come in different variants~\cite{cone}. They have in common that
the jet is defined by the particles found inside a circle in the plane formed by rapidity 
and azimuthal angle, such that the sum of the four-momenta of these particles
points in the direction of its center. While widely used in experiment, the traditional
cone algorithms (notably the ones known as ``midpoint'' algorithm mostly used at 
RHIC~\cite{star,star1} and the ``iterative cone'' algorithm) 
were found to be not infrared-safe~\cite{irsafe,soyez1}. This evidently sets a serious 
limitation to the use of such algorithms in the theoretical calculation and to comparisons 
of data and theory. For single-inclusive jet cross sections, the lack of infrared-safety 
becomes an issue first at next-to-next-to-leading order (NNLO) in perturbation theory, so 
that next-to-leading order (NLO) calculations remain meaningful in the sense that they
produce finite and well-defined answers. In case of the midpoint cone algorithm,
a solution to the problem of infrared-unsafety was found in terms of the 
``Seedless Infrared Safe Cone'' (SISCone) algorithm~\cite{soyez1}. It was
also shown that the anti-$k_t$ algorithm mentioned above can effectively cure
the lack of infrared-safety of the iterative cone algorithm~\cite{css}. As a result,
the SISCone and all $k_t$-type algorithms are nowadays known to be
infrared-safe and are preferred for use in experiments. 

In earlier work~\cite{jet1,jet2}, the spin-dependent (and spin-averaged) cross sections
for $pp\to {\mathrm{jet}}\,X$ were derived at NLO.  Ref.~\cite{jet1} was based on a 
Monte-Carlo integration approach, while~\cite{jet2} used a largely analytic technique 
for deriving the relevant partonic cross sections for cone algorithms, 
which becomes possible if one
assumes the jet to be a rather narrow object~\cite{jet2,oldsca,oldsca1,oldsca2,oldsca3}. 
This assumption is equivalent to the approximation that the cone opening $R$ of the jet 
is not too large, and hence was termed ``Small Cone Approximation'' (SCA) in~\cite{jet2}. 
In the SCA, one systematically expands the partonic cross sections around $R=0$. 
The dependence on $R$ is of the form ${\cal A}\log R+{\cal B}+{\cal O} (R^2)$. 
The coefficients ${\cal A}$ and ${\cal B}$ are retained and calculated analytically,
whereas the remaining terms ${\cal O} (R^2)$ and beyond are neglected. 
The advantage of the analytical method is that it leads to much faster and more 
efficient computer codes and is hence readily suited for inclusion of jet spin asymmetry 
data from RHIC in a NLO global analysis of polarized parton distributions. Indeed, the 
results of~\cite{jet2} have been used in the global analysis~\cite{dssv}, where the 
experimental data from STAR~\cite{star} were used to constrain $\Delta g$. 
It was shown in~\cite{jet2} that the SCA is in fact an excellent approximation to the 
full Monte-Carlo calculation (which is valid for arbitrary cone openings), for all values of $R$ and
kinematics relevant at RHIC. (We shall revisit this finding in our phenomenological 
section~\ref{Pheno} below).

Commensurate with the procedure chosen by STAR, the calculation~\cite{jet2} 
was performed for the midpoint cone algorithm. 
%We recall that the infrared-unsafety 
%of that algorithm is not an issue for the single-inclusive jet cross section at NLO. 
In the present paper, we will extend the work in~\cite{jet2} to the case
of  $k_t$-type algorithms. We will again use the approximation of a rather
narrow jet. As we do not really have a jet ``cone'' for the $k_t$-type algorithms,
we shall from now on refer to this approximation as ``Narrow Jet Approximation'' (NJA). 
This term will be collectively applied to both the cone (where it used to be the SCA)
and the $k_t$-type algorithms. The meaning of the ``NJA'' will always be that the 
jet parameter $R$ used to define the jet (cone opening for the cone algorithm and
distance between two objects for the $k_t$-type algorithms) is not too large, as we shall 
discuss in more detail below. 

One motivation for our new study is that $k_t$-type algorithms 
are also being considered in STAR now~\cite{staralg}, so that it is timely
to prepare the corresponding theoretical NLO calculations for the spin asymmetries. 
The differences between the jet cross sections for the cone and $k_t$-type algorithms 
in the NJA are also interesting from a theoretical point of view. We will find
that they amount to finite contributions with leading order (LO) kinematics.
As such they play a role as matching coefficients in threshold resummation studies of 
jet production, as was discussed in~\cite{ddfwv}. They also appear in a related 
context in studies of jet shapes in the framework of  ``Soft Collinear Effective
Theories'' (SCET)~\cite{scet}. 

The remainder of this paper is organized as follows: In Sec.~\ref{tech} we
present the technical details and analytical results of our calculation of single-inclusive
jet cross sections in the NJA, focusing on the $k_t$-type algorithms. 
Section~\ref{Pheno} contains phenomenological results relevant for RHIC.
We summarize our work in Sec.~\ref{sum}.

%%%%%%%%%%%%%%%%%%%%%%%%%%%%%%%%%%%%%%%%%
\section{Technical Details \label{tech}}
%%%%%%%%%%%%%%%%%%%%%%%%%%%%%%%%%%%%%%%%%
\subsection{Cone and $k_t$-type jet definitions \label{jetdef}}
%%%%%%%%%%%%%%%%%%%%%%%%%%%%%%%%%%%%%%%%%

We consider single-inclusive jet production in hadronic collisions, 
$pp \rightarrow {\mathrm{jet}}\,X$, where the jet has a transverse momentum $p_{T_J}$,
pseudorapidity $\eta_J$, and azimuthal angle $\phi_J$.  
The cross section is infinite unless a finite jet size is imposed as 
a parameter. The different jet algorithms vary in the way this size is defined. In the cone 
algorithm~\cite{cone}, one defines the jet by all particles $j$ that satisfy 
\begin{equation}
\label{eq:conedef}
R_{jJ}^2\equiv (\eta_J-\eta_j)^2 + (\phi_J-\phi_j)^2 \le R^2  .
\end{equation}
Here $\eta_j$ and $\phi_j$ denote the pseudo-rapidities and azimuthal
angles of the particle, and $R$ is the jet cone aperture. The jet four-momentum 
sets the center of the cone; it is nowadays usually defined as the sum of the
four-momenta of the particles $j$ forming the jet. 

For the $k_t$-type algorithms~\cite{catani,catani1,CA,css} 
one defines for each pair of objects (initially, particles) 
$j,k$ the quantity
\ba \label{dist1}
d_{jk}\equiv {\mathrm{min}} ( k_{T_j}^{2 p}, k_{T_k}^{2 p}) {R_{jk}^2\over R^2},
\ea
where $p$ is a parameter that specifies the algorithm, $k_{T_j}$ denotes the transverse 
momentum of particle $j$ with respect to the beam direction, and 
\ba\label{Rij}
R_{jk}^2\equiv (\eta_j-\eta_k)^2 + (\phi_j-\phi_k)^2.
\ea
The parameter $R$ is called the jet radius. $d_{jk}$ may be viewed as a 
distance between two objects $j$ and $k$. One also defines for each object
a distance to the initial beams:
\ba\label{dist2}
d_{jB}\equiv k_{T_j}^{2 p}. 
\ea
The algorithm identifies the smallest of the $d_{jk}$ and $d_{jB}$. If it is a beam distance,
the object is defined as a jet and removed from the list of objects.  
If the smallest distance is a $d_{jk}$, the two objects $j,k$ are
merged into a single one. The procedure is 
repeated until no objects are left in the event. As mentioned above, the jet algorithm
is fully specified by the parameter $p$. We have $p=1$ for the $k_t$-algorithm~\cite{catani,catani1}, 
$p=0$ for the Cambridge/Aachen algorithm~\cite{CA}, and $p=-1$ for the 
anti-$k_t$ algorithm~\cite{css}. 

Note that on top of the choice of jet algorithm one also has to define how objects
are to be merged if the need for that arises. Throughout this paper our choice will be 
(for both algorithms) to define the four-momentum of a new object as the sum 
of four-momenta of the partons that form the new object. This so-called ``$E$ recombination 
scheme''~\cite{cone} is the most popular choice nowadays.

%%%%%%%%%%%%%%%%%%%%%%%%%%%%%%%%%%%%%%%%%%%%%%%%%%%%%%
\subsection{Calculation of single-inclusive jet production cross sections at NLO}
%%%%%%%%%%%%%%%%%%%%%%%%%%%%%%%%%%%%%%%%%%%%%%%%%%%%%%

The spin-averaged cross section for the process  $p(P_a)p(P_b)\rightarrow\mathrm{jet}(p_J)\,X$ 
can be written as~\cite{jet2}
\begin{eqnarray} 
\frac{d^2 \sigma}{dp_{T_J}d\eta_J} 
&=& \frac{2 p_{T_J}}{S} \sum_{a,b} 
\int_{VW}^{V}\frac{dv}{v(1-v)}  \nonumber\\[1mm]&&\hspace*{-5mm} 
\times \int^1_{VW/v}\!\!\frac{dw}{w}f_a(x_a,\mu_F) f_b(x_b,\mu_F) 
\nonumber \\[1mm]
&&\hspace*{-5mm} \times \Big [ 
\frac{d \hat{\sigma}^{(0)}_{ab\rightarrow \mathrm{jet} X}(s,v)}{dv} 
\delta (1-w)  + \nonumber\\[1mm]&& \hspace*{-5mm}  \frac{\alpha_s(\mu_R)}{\pi} \, \frac{d^2
\hat{\sigma}^{(1)}_{ab\rightarrow \mathrm{jet}X}(s,v,w,\mu_F,\mu_R; R)}
{dvdw} \Big ] \label{jetcross} , 
\end{eqnarray}
where the dimensionless variables $V$ and $W$ are defined in terms of $p_{T_J}$ and $\eta_J$ as
\begin{equation}
V=1-\frac{p_{T_J}}{\sqrt{S}}\; e^{\eta_J}\;\;\;\mathrm{and}\;\;\;
W=\frac{p_{T_J}^2}{S V (1-V)},
\label{VW}
\end{equation}
with $\sqrt{S}=\sqrt{(P_a+P_b)^2}$ the hadronic c.m.s.\ energy.
$v$ and $w$ are the corresponding parton-level variables; they
are given in terms of the partonic Mandelstam variables
\be
s\equiv (p_a+p_b)^2,\;\;t\equiv (p_a-p_J)^2,\;\;u\equiv(p_b-p_J)^2,
\ee
as 
\ba
v= 1+\frac{t}{s},\;\; w= \frac{-u}{s+t}.
\label{partvw}
\ea
The sum in~(\ref{jetcross}) runs over all partonic channels $a+b \to \mathrm{jet}+X$, with 
$d \hat{\sigma}^{(0)}_{ab\rightarrow \mathrm{jet} X}$ and $d \hat{\sigma}^{(1)}_{ab\rightarrow \mathrm{jet} X}$ 
the LO and NLO terms in the corresponding partonic cross sections, respectively. $f_a(x_a,\mu_F)$
and $f_b(x_b,\mu_F)$ denote the parton distribution functions at factorization scale $\mu_F$ whose 
partonic momentum fractions are determined by $V,W,v$ and $w$:
\begin{equation}
\label{xaxb}
x_a = \frac{V W}{v w},\;\;\;
x_b=  \frac{1-V}{1-v}.
\end{equation}
Finally, $\mu_R$ in~(\ref{jetcross}) is the renormalization scale for the strong coupling constant.

We note that expression~(\ref{jetcross}) can be straightforwardly extended to the case of collisions of
longitudinally polarized protons. Here one defines a spin-dependent cross section as
\be\label{heldef}
\frac{d^2 \Delta \sigma}{dp_{T_J}d\eta_J} \equiv \frac{1}{2} \left[
\frac{d^2 \sigma^{++}}{dp_{T_J}d\eta_J}-\frac{d^2 \sigma^{+-}}{dp_{T_J}d\eta_J}\right],
\ee
where the superscripts indicate the helicities of the colliding protons. The structure of Eq.~(\ref{jetcross})
also applies to $d^2 \Delta \sigma/dp_{T_J}d\eta_J$, except that the partonic cross sections and parton
distribution functions are to be replaced by their spin-dependent counterparts 
$d \Delta \hat{\sigma}_{ab\rightarrow \mathrm{jet} X}$ and $\Delta f_{a,b}(x_{a,b},\mu_F)$, 
respectively. The former are defined in analogy with~(\ref{heldef}), and for the latter we have
\be
\Delta f_a(x,\mu_F)=f_a^+(x,\mu_F) - f_a^-(x,\mu_F),
\ee
where $f_a^+$ ($f_a^-$) denotes the distribution for partons of type $a$ with the same (opposite)
helicity as that of the parent proton. All our expressions below will be formulated for the spin-averaged
case; however, they equally apply to the polarized case with the modifications just discussed.

A possible way of organizing the NLO calculation of the single-inclusive jet cross section
was developed and employed in Refs.~\cite{jet2,oldsca1,oldsca2,oldsca3,Guillet}. It starts from the NLO 
single-parton inclusive cross sections $d\hat\sigma_{ab\rightarrow c X}$, relevant for the 
single-inclusive hadron production process $pp\to hX$ and analytically known from previous
calculations~\cite{oldsca2,jssv}.  These cross sections cannot directly 
be used to describe jet production; they can, however, be converted to the desired single-inclusive 
jet cross sections. To this end, one first imagines a ``jet cone'' around the observed parton 
$c$ and notices that a NLO single-parton inclusive cross section contains configurations where there is 
an additional parton $d$ inside the cone (note that we use the term ``cone'' here
just for simplicity-- the considerations apply to any jet definition). 
For a jet cross section, the observed final state should 
not just be given by parton $c$, but by partons $c$ and $d$ jointly. One therefore subtracts these 
contributions and replaces them by terms for which partons $c$ and $d$ are both inside the cone and 
form the observed jet together. To be more precise, for a given partonic process $ab\to cde$ we have,
after proper bookkeeping of all partonic configurations that are possible in the cone:
\begin{eqnarray}
d \hat{\sigma}_{ab\rightarrow \mathrm{jet}X} &=&  
[d \hat{\sigma}_c -d \hat{\sigma}_{c(d)}-
d \hat{\sigma}_{c(e)}]\nonumber\\[1mm] &+&
 [d \hat{\sigma}_d -d \hat{\sigma}_{d(c)}-
d \hat{\sigma}_{d(e)}]\nonumber\\[1mm] &+&
[d  \hat{\sigma}_e -d \hat{\sigma}_{e(c)}-
d \hat{\sigma}_{e(d)}]\nonumber\\[1mm] &+&
d \hat{\sigma}_{cd} + d \hat{\sigma}_{ce}+ 
d \hat{\sigma}_{de} .
\label{jetform}
\end{eqnarray}
Here $d \hat{\sigma}_j$ is the single-parton inclusive cross section where parton $j$ 
is observed (which also includes the virtual corrections),  $d \hat{\sigma}_{j(k)}$ is the cross section 
where parton $j$ is observed but parton $k$ is also in the cone, and $d \hat{\sigma}_{jk}$ is the cross 
section when both partons $j$ and $k$ are inside the cone and jointly form the jet. 

The single-parton inclusive cross sections $d \hat{\sigma}_j$ of~\cite{oldsca2,jssv} 
were obtained after a subtraction of final-state collinear singularities in the 
modified minimal subtraction ($\overline{\mathrm{MS}}$) scheme. Upon calculation of 
the combinations $d \hat{\sigma}_{j(k)}+d \hat{\sigma}_{k(j)}-d \hat{\sigma}_{jk}$ 
in~(\ref{jetform}) one also finds collinear singularities, which must match those initially
present in $(d \hat{\sigma}_j+d \hat{\sigma}_k)/2$. On the other hand, the full expression
in Eq.~(\ref{jetform}), being an inclusive-jet cross section, must be collinear-finite.
Therefore, in order to obtain the combination in~(\ref{jetform}) correctly, one just needs 
to perform an $\overline{\mathrm{MS}}$ subtraction also of the singularities in the 
$d \hat{\sigma}_{j(k)}+d \hat{\sigma}_{k(j)}-d \hat{\sigma}_{jk}$. For further discussion, see~\cite{jet2}.

In practice, it is convenient to consider the $d \hat{\sigma}_{j(k)}$ and $d \hat{\sigma}_{jk}$
separately. In the NJA, to which we will turn in the next subsection, they may in fact be computed 
analytically. At NLO, they both receive contributions only from real-emission $2\to 3$ diagrams. 
Since for $d \hat{\sigma}_{j(k)}$ the jet is obtained from the single parton $j$, it is
independent of the jet definition. The relevant results for this piece in the NJA may be found in~\cite{jet2}.
For $d \hat{\sigma}_{jk}$, on the other hand, the situation is different since here both particles 
$j$ and $k$ jointly form the jet. In Refs.~\cite{jet2,oldsca1,oldsca3} the $d \hat{\sigma}_{jk}$ were 
obtained for the case of cone algorithms. In the following we will derive them also for the 
$k_t$-type algorithms, using again the NJA.

%%%%%%%%%%%%%%%%%%%%%%%%%%%%%%%%%%%%%%%%%%%%%%%%%%
\subsection{Calculation of $d \hat{\sigma}_{jk}$ in the ``Narrow Jet Approximation'' \label{NJA}}
%%%%%%%%%%%%%%%%%%%%%%%%%%%%%%%%%%%%%%%%%%%%%%%%%%

It is instructive to discuss the cone and $k_t$-type algorithms in parallel, in order to make
contact with the derivations made in~\cite{jet2} and to make our paper self-contained. 
From Sec.~\ref{jetdef}, we see that both the 
cone and the $k_t$-type algorithms contain a jet parameter $R$. In the NJA one assumes 
$R$ to be relatively small. As discussed at the end of Sec.~\ref{jetdef}, our choice
is to merge objects by adding their four-momenta. For $d \hat{\sigma}_{jk}$ this means that 
the four-momentum $p_J$ of the jet is the sum of the parton four-momenta $p_j$ and $p_k$. 

We first observe that for
the $k_t$-type algorithms the two partons $j,k$ are merged into one jet if their distance
defined in~(\ref{dist1}) is smaller than their respective beam distances $d_{iB}$ and $d_{jB}$ 
defined in~(\ref{dist2}). For $d \hat{\sigma}_{jk}$ this has to be the case by definition, and we
therefore arrive at the condition
\be\label{j1}
R_{jk}^2\leq R^2 \quad \quad {\mathrm{for}}\; k_t{\mathrm{-type \;algorithms}},
\ee
with $R_{ij}$ defined in Eq.~(\ref{Rij}). We stress that this condition holds for {\it all}
$k_t$-type algorithms, regardless of the choice of the parameter $p$. This implies
that at NLO all $k_t$-type algorithms lead to the same jet cross section, a result
that does not rely on the NJA. Equation~(\ref{j1}) is to be contrasted with 
Eq.~(\ref{eq:conedef}) valid for the cone algorithm:
\be\label{j2}
R_{jJ}^2 \leq R^2\;  \wedge\; R_{kJ}^2 \leq R^2 \;\;
\quad {\mathrm{for\; cone \;algorithm}}.
\ee
The difference between Eqs.~(\ref{j1}) and~(\ref{j2}) is solely responsible for any
differences between the NLO results for the two types of jet algorithms! For
the $k_t$-type algorithms it is the ``distance'' between the two partons that 
is constrained by the jet algorithm, whereas for the cone algorithm it is 
the distance of each parton to the jet itself. Note that this observation was
already made in Ref.~\cite{catani}, and in~\cite{scet} in a slightly different context. 

As was shown in~\cite{jet2}, in the NJA $d \hat{\sigma}_{jk}$
is, up to trivial factors, given by the following expression 
(see Eqs.~(19),(20),(27) of that paper):
\be
\label{eq:scaps3}
\frac{d \hat{\sigma}_{jk}}{dvdw}\,\propto\,\int \frac{dPS_3}{dvdw}\, 
\frac{P_{jK}^{<}(z)}{2p_j\cdot p_k},
\ee
where the integration $dPS_3$ is over the phase space of the three-body
final state of the overall partonic process, which is carried out in 
$d=4-2\varepsilon$ dimensions. 
Expression~(\ref{eq:scaps3}) arises from the fact that $d \hat{\sigma}_{jk}$ is 
strongly dominated by contributions for which particles $j$ and $k$ result from  
collinear splitting of an intermediate particle $K$. The reason is that
in this case the propagator of the intermediate particle, represented by the
denominator $1/(2p_j\cdot p_k)$, goes on-shell. For instance,  
if the jet is formed by a quark and a gluon, the pair will predominantly 
originate from a quark splitting into a quark plus a gluon, described by the 
splitting functions $P_{qq}$ and $P_{gq}$. The argument $z$ of the splitting
function is the fraction of the intermediate particle's momentum transferred in
the splitting. The superscript ``$<$'' on the splitting function indicates that the 
$d$-dimensional splitting function $P_{jK}(z)$ is strictly at $z<1$, that is, without 
its $\delta(1-z)$ contribution that is present when $j=K$. This is a necessary condition 
in order to have two partons producing the jet. For additional details, we refer the 
reader to Ref.~\cite{jet2}.

Making use of the fact that $p_J=p_j+p_k$, the term on the right-hand-side
of~(\ref{eq:scaps3}) may be written as~\cite{jet2}
\begin{eqnarray}
\label{eq:scaps4}
\int \frac{dPS_3}{dvdw} \frac{P_{jK}^{<}(z)}{(2p_j\cdot p_k)} &=&
\left[ \frac{1}{8\pi} \left( \frac{4\pi}{s}
\right)^{\varepsilon}\frac{(v(1-v))^{-\varepsilon}}{\Gamma(1-\varepsilon)}
\right]\nonumber\\[2mm]
&&\hspace*{-3.cm}\times \frac{1}{8\pi^2}\left( \frac{4\pi}{s}
\right)^{\varepsilon} \frac{\delta(1-w)}{\Gamma(1-\varepsilon)} 
\int_0^{E_J} dE_j \,\frac{E_J}{E_k^2}
\left( \frac{E_j^2}{s}\right)^{-\varepsilon}\nonumber \\[2mm]
&&\hspace*{-3.cm}\times P_{jK}^{<}(z)\,
\int_0^{\theta_{\mathrm{max}}} d\theta_{j}\frac{\sin^{1-2 \varepsilon}
\theta_{j}}{1-\cos\theta_{jk}} ,
\end{eqnarray}
where $E_J=E_j+E_k$ is the jet energy (with $E_{j,k}$ the energies of 
partons $j,k$), $z=E_j/E_J$, and $\theta_{jk}$ the 
angle between the three-momenta of partons $j$ and $k$. $\theta_j$ is the polar 
angle of parton $j$, measured with respect to the jet direction. $\theta_{\mathrm{max}}$
is an upper limit on the $\theta_j$-integration that needs to be derived according to
the jet algorithm. It is of the order of the jet parameter $R$, and hence treated
as small in the NJA.

It is useful to write the $\theta_j$-integral as an integral over the 
(squared) invariant mass $p_J^2\equiv m^2=2p_j\cdot p_k$ of the produced jet. 
One finds
\begin{eqnarray}
\cos (\theta_{jk})&=& 1-\frac{m^2}{2 E_j E_k},\nonumber \\[1mm]
\cos (\theta_j) &=& \frac{2 E_j E_J - m^2}{2 E_j \sqrt{E_J^2 - m^2}}.
\end{eqnarray}
With this one obtains after some algebra:
\begin{eqnarray}
\label{eq:scaps5}
\int \frac{dPS_3}{dvdw} \frac{P_{jK}^{<}(z)}{(2p_j\cdot p_k)} &=&
\left[ \frac{1}{64\pi^3} \left( \frac{4\pi}{\sqrt{s}}
\right)^{2\varepsilon}\frac{(v(1-v))^{-\varepsilon}}{\Gamma^2(1-\varepsilon)}
\right]\nonumber\\[2mm]
&&\hspace*{-3.5cm}\times \delta(1-w)
\int_0^1 dz\, z^{-\varepsilon} (1-z)^{-\varepsilon}P_{jK}^{<}(z)
\,\int_0^{m^2_{{\mathrm{max}}}} \frac{dm^2}{m^2} m^{-2 \varepsilon},
\nonumber \\
\end{eqnarray}
where we have also expressed the integral over the energy $E_j$ as an 
integral over $z$. In the NJA, $m^2$ is a small quantity, and we have hence
neglected powers of $m^2$ wherever possible. Note however that 
the $m^2$-integral produces a $1/\varepsilon$-singularity at the lower end. 

All that is left to be done now is to determine the upper limit of the integral
over $m^2$, which depends on the jet algorithm chosen. In order to make
contact with the results in Ref.~\cite{jet2}, we do this first for the cone
algorithms and afterwards for the $k_t$-type algorithms we are mainly
interested in here. In both cases we first write the jet four-momentum as 
\be
p_J = \left( E_J,|\vec{p}_J|\frac{\cos(\phi_J)}{\cosh(\eta_J)},
|\vec{p}_J|\frac{\sin(\phi_J)}{\cosh(\eta_J)},|\vec{p}_J|\tanh(\eta_J)
\right),
\ee
where
\be
|\vec{p}_J|=\sqrt{E_J^2-m^2}.
\ee
For the parton momenta $p_j$ and $p_k$, which are light-like, we write accordingly
\ba\label{p1}
p_j &=& E_j\left(1,\frac{\cos(\phi_j)}{\cosh(\eta_j)},
\frac{\sin(\phi_j)}{\cosh(\eta_j)},\tanh(\eta_j)\right),\nn\\[2mm]
p_k &=& E_k\left(1,\frac{\cos(\phi_k)}{\cosh(\eta_k)},
\frac{\sin(\phi_k)}{\cosh(\eta_k)},\tanh(\eta_k)\right),
\ea
with the azimuthal angles $\phi_{j,k}$ and pseudorapidities $\eta_{j,k}$ 
of the partons. \\
{\it (i) Cone algorithms:} 
We write
\ba\label{long}
m^2& =& 2 p_j \cdot p_k = 2 p_j \cdot p_J\nonumber \\[1mm]
&&\hspace*{-1cm}\approx 2 E_j E_J\frac{
\cosh(\eta_j-\eta_J)-\cos(\phi_j-\phi_J)}{\cosh(\eta_j)\cosh(\eta_J)}\nn\\[2mm]
&&\hspace*{-1cm}+ \frac{E_j m^2}{E_J}\frac{
\cos(\phi_j-\phi_J)+\sinh(\eta_j)\sinh(\eta_J)}{\cosh(\eta_j)\cosh(\eta_J)},
\ea
where we have expanded $|\vec{p}_J|$ to first order in $m^2$.
The combination $\left(\cosh(\eta_j-\eta_J)-\cos(\phi_j-\phi_J)\right)$ in~(\ref{long})
is small, while in the other term $m^2$ is small. Therefore, to the order we consider, 
we can set $\eta_j=\eta_J$, $\phi_j=\phi_J$ in all other places. This gives
\ba
m^2 \approx \frac{E_j E_J}{\cosh^2(\eta_J)}\,R_{jJ}^2 
+ \frac{E_j m^2}{E_J},
\ea
where $R_{jJ}$ is as defined in Eq.~(\ref{eq:conedef}). 
We solve for $m^2$ and get
\ba
m^2 \approx  \frac{E_J^2}{\cosh^2(\eta_J)}\,\frac{z}{1-z}\,R_{jJ}^2.
\ea
Likewise, we find
\be
m^2 \approx   \frac{E_J^2}{\cosh^2(\eta_J)}\,\frac{1-z}{z}\,R_{kJ}^2.
\ee
The jet criterion~(\ref{j2}) in the cone algorithm then immediately
translates into
\be
m^2_{{\mathrm{max,cone}}} = \frac{E_J^2R^2}{\cosh^2(\eta_J)} 
\min\left( \frac{z}{1-z},\frac{1-z}{z}\right).
\ee
The last two integrals in~(\ref{eq:scaps5}) are now readily performed:
\ba\label{Ijk1}
&&\hspace*{-9mm}
\int_0^1 dz\, z^{-\varepsilon} (1-z)^{-\varepsilon}P_{jK}^{<}(z)
\,\int_0^{m^2_{{\mathrm{max,cone}}}} \frac{dm^2}{m^2} m^{-2 \varepsilon}\nn\\[2mm]
&=&-\frac{1}{\varepsilon}\,\left( \frac{E_J^2R^2}{\cosh^2(\eta_J)} \right)^{-\varepsilon}
I_{jK}^{{\mathrm{cone}}},
\ea
where
\be
I_{jK}^{{\mathrm{cone}}}\equiv
\left[\int_0^{1/2} dz  z^{-2 \varepsilon}+\int_{1/2}^1 dz (1-z)^{-2 \varepsilon}\right]P_{jK}^{<}(z).
\ee
The explicit results for these integrals for the various splitting functions 
were given in~\cite{jet2}:
\begin{eqnarray}
I_{qq}^{{\mathrm{cone}}}&=& C_F \left[ -\frac{1}{\varepsilon} -\frac{3}{2} +\varepsilon
\left(  -\frac{7}{2}+\frac{\pi^2}{3} -3 \log 2 \right) \right]\nonumber \\[2mm]
&=&I_{gq}^{{\mathrm{cone}}}, \nonumber \\[2mm]
I_{qg}^{{\mathrm{cone}}}&=& \frac{1}{2} \left[\frac{2}{3} +\varepsilon
\left(  \frac{23}{18}+\frac{4}{3}\log 2 \right) \right]  , \nonumber \\[2mm]
I_{gg}^{{\mathrm{cone}}}&=&2 C_A \left[ -\frac{1}{\varepsilon} -\frac{11}{6} +\varepsilon
\left( -\frac{137}{36} +\frac{\pi^2}{3} -\frac{11}{3} \log 2 \right) \right],
\nn\\
\label{eq:imn}
\end{eqnarray}
with $C_A=3$ and $C_F=4/3$ the usual SU(3) Casimir operators.
Note that the ratio $R/\cosh(\eta_J)$ corresponds to the half-opening 
$\delta$ of the jet cone considered in~\cite{jet2}. We also note the
logarithmic dependence of $d \hat{\sigma}_{jk}$ on $R$ arising from the 
factor $R^{-2\varepsilon}$ in Eq.~(\ref{Ijk1}). \\
{\it (ii) $k_t$-type algorithms:} 
Here we use Eq.~(\ref{p1}) to directly compute
\be
m^2 =\frac{2E_j E_k}{\cosh(\eta_j)\cosh(\eta_k)} \left(\cosh(\eta_j-\eta_k)-
\cos(\phi_j-\phi_k) \right).
\ee
We can approximate this by
\ba
m^2  &\approx&\frac{E_j E_k}{\cosh^2(\eta_J)} \left((\eta_j-\eta_k)^2+
(\phi_j-\phi_k)^2 \right)\,\nonumber\\[2mm]&&
=\frac{E_j E_k}{\cosh^2(\eta_J)} R_{jk}^2,
\ea
with $R_{jk}$ defined in~(\ref{j1}). The condition $R_{jk}^2\leq R^2$ then 
immediately gives
\be
m^2_{{\mathrm{max,}}k_t}= \frac{E_J^2R^2}{\cosh^2(\eta_J)}   z (1-z),
\ee
and instead of~(\ref{Ijk1}) we have
\ba\label{xint1}
&&\hspace*{-9mm}
\int_0^1 dz\, z^{-\varepsilon} (1-z)^{-\varepsilon}P_{jK}^{<}(z)
\,\int_0^{m^2_{{\mathrm{max}},k_t}} \frac{dm^2}{m^2} m^{-2 \varepsilon}\nn\\[2mm]
&=&-\frac{1}{\varepsilon}\,\left( \frac{E_J^2R^2}{\cosh^2(\eta_J)} \right)^{-\varepsilon}
I_{jK}^{k_t},
\ea
where
\ba\label{xint2}
I_{qq}^{k_t}&=&C_F \left[ -\frac{1}{\varepsilon}-\frac{3}{2}+\varepsilon\left(
-\frac{13}{2}+\frac{2\pi^2}{3}\right)
\right]\nn\\[2mm]
&=&I_{gq}^{k_t},\nn\\[2mm]
I_{qg}^{k_t}&=&\frac{1}{2}\left[ \frac{2}{3}+\frac{23}{9}\varepsilon \right],\nn\\[2mm]
I_{gg}^{k_t}&=&2C_A \left[ -\frac{1}{\varepsilon}-\frac{11}{6}+\varepsilon\left(-\frac{67}{9}+
\frac{2\pi^2}{3}\right)\right].
\ea
Comparison with~(\ref{eq:imn}) shows that the pole terms 
in Eqs.~(\ref{Ijk1}),(\ref{xint1}) are the same, 
as they have to be. The finite remainders, however,
differ and will lead to finite and $R$-independent 
differences in the NLO cross sections for the two
types of algorithms. As seen from Eq.~(\ref{eq:scaps4}), the cross sections
$d\hat{\sigma}_{jk}$ are proportional to $\delta(1-w)$ and hence have LO
kinematics. This will, therefore, also be true for the finite differences
just mentioned.
We note that expressions similar to those in~(\ref{xint2}) were also obtained
in~\cite{scet} in the context of jet studies in SCET. 
%%%%%%%%%%%%%%%%%%%%%%%%%%%%%%%%%%%%%%%%%%%%%%%%
\begin{figure*}[t!]
\centering
\includegraphics[width=8.5cm,clip]{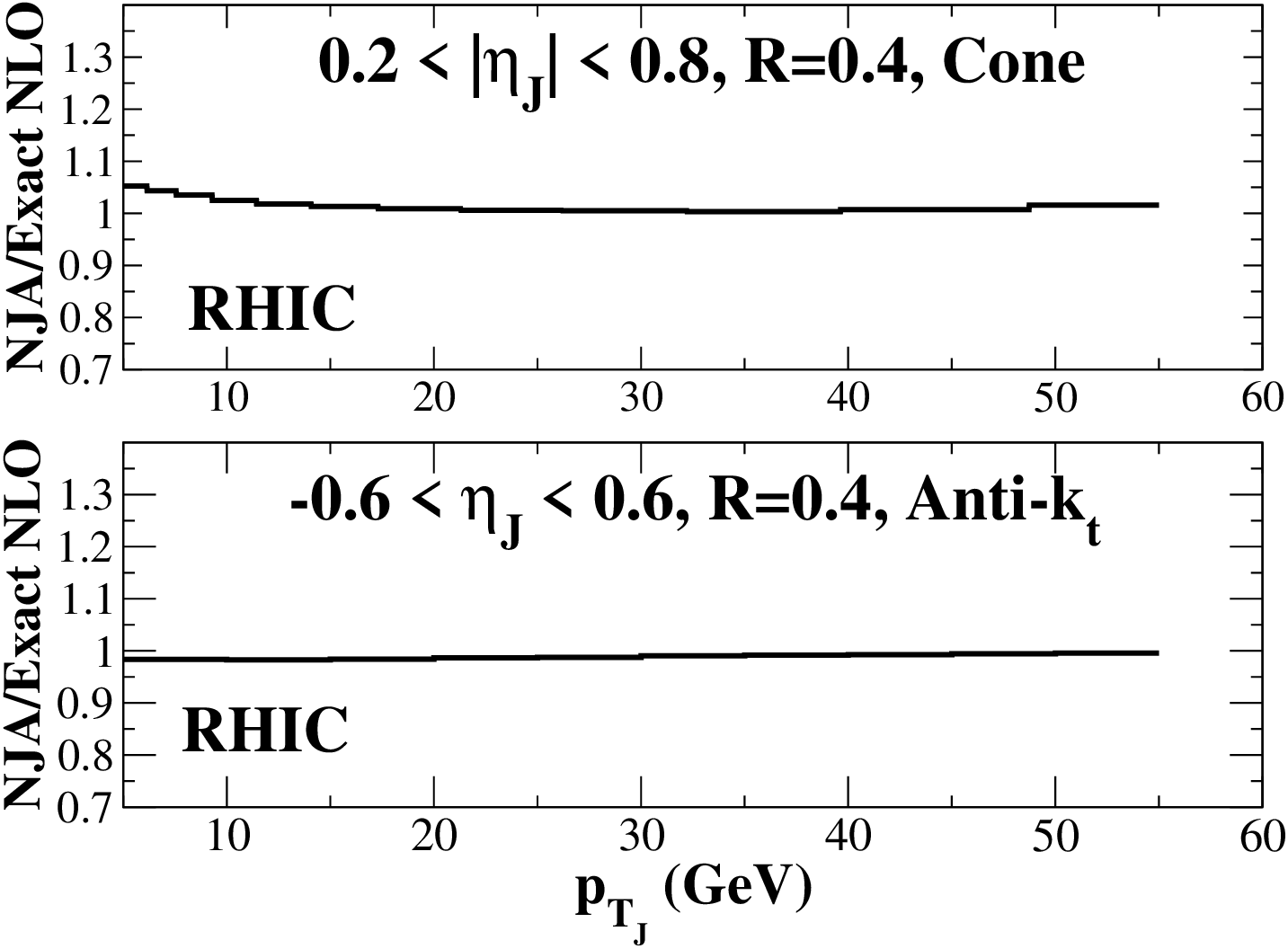}
\hspace{0.2cm}
\includegraphics[width=8.7cm,clip]{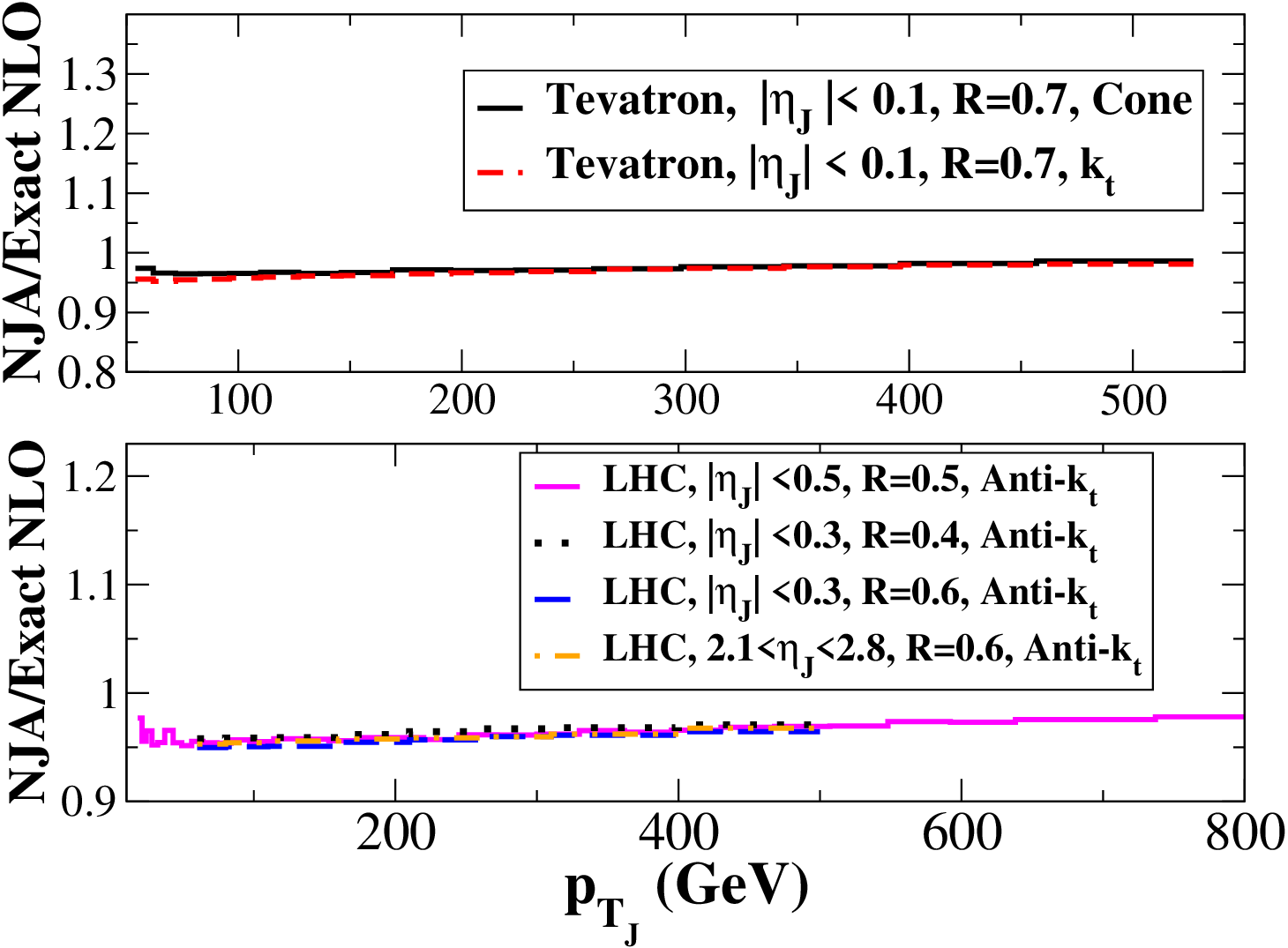}
\caption{\label{fig1}
\sf Upper left: Ratio of single-inclusive jet cross sections at RHIC for the cone
algorithm, as computed within the NJA and with 
{\it fastNLO}~\cite{fastnlo}. Lower left: Same for the jet cross sections for
the $k_t$-type algorithms. Here, the exact NLO calculation was
performed with the {\it FastJet} code~\cite{Cacciari:2011ma,greg}. 
Right: Similar comparisons for Tevatron (upper, $\sqrt{S}=1960$~GeV) and
LHC (lower, $\sqrt{S}=7$~TeV) energies. 
The exact NLO results for Tevatron and for LHC with $R=0.5$ were obtained
from {\it fastNLO}, the others from {\it FastJet}.
}
\end{figure*}  
%%%%%%%%%%%%%%%%%%%%%%%%%%%%%%%%%%%%%%%%%%%%%%%%

It is now straightforward to use the integrals $I_{jK}^{k_t}$ given above
to derive the NLO jet cross section for the $k_t$-type algorithms in the NJA--
the calculation proceeds exactly as in~\cite{jet2}. In fact, is is very easy to 
change the numerical code of~\cite{jet2} to the case of the $k_t$-type algorithms:
The integrals $I_{jK}^{{\mathrm{cone}}}$ are the only sources of terms
$\propto \log 2$ in the NLO calculation for the cone algorithms. Thus, by 
replacing these terms appropriately in each subprocess one can translate
the result from the cone algorithm to the case of $k_t$-type algorithms
without having to do the full calculation in~(\ref{jetform}).

While we have derived all results above for the spin-averaged cross section, it 
is straightforward to extend them to the case of jet production in polarized collisions. 
In the NJA, the contributions by particles $j$ and $k$ forming the jet entirely 
arise from {\it final-state} emission, which is independent of the polarization of the
initial partons. Therefore the same integrals $I_{jK}^{{\mathrm{cone}}}$ or
$I_{jK}^{k_t}$ apply to the polarized case.

%%%%%%%%%%%%%%%%%%%%%%%%%%%%%%%%%%%%%%%%%%%%%%%%
\section{Phenomenological Results \label{Pheno}}
%%%%%%%%%%%%%%%%%%%%%%%%%%%%%%%%%%%%%%%%%%%%%%%%

In this section, we present some phenomenological results for single-inclusive jet 
production cross sections and spin asymmetries in $pp$ collisions at RHIC. 
Our main focus is of course on the sensitivity of these quantities 
to the jet algorithm used. 

\subsection{Unpolarized collisions}

We begin by ascertaining the accuracy of the NJA. It was shown already in~\cite{jet2}
that for the cone algorithm the approximation is very accurate for the values of $R$ and 
kinematics relevant at RHIC. To confirm this finding, we make use of the 
recently developed {\it fastNLO} package~\cite{fastnlo} which is based on the NLO
code {\it NLOJet++} of~\cite{Nagy} and allows to compute NLO jet
cross sections with Monte-Carlo integration methods. In fact, the authors of the code
offer an online tool that provides numbers for the jet cross section at RHIC
for the mid-point cone algorithm, as used by STAR~\cite{star}. In the upper left part of
Fig.~\ref{fig1} we compare these results to the ones we find with our code based
on the NJA. We plot the ratio of the two theoretical results.
We have used here the CTEQ6.6M parton distributions~\cite{cteq66}, which will be 
our choice for the spin-averaged parton distribution functions throughout this 
paper. The comparison is for $\sqrt{S}=200$~GeV, $R=0.4$, and a range of 
pseudorapidity $0.2\leq |\eta_J|\leq 0.8$. We have chosen the factorization and 
renormalization scales as $\mu_F=\mu_R=p_{T_J}$. As one can see, there is 
excellent agreement between the full NLO Monte-Carlo calculation and our 
approximated result. The largest deviations occur at the lowest $p_{T_J}$; even 
here they amount to at most $5\%$. 

The lower left part of Fig.~\ref{fig1} shows a similar comparison for the 
$k_t$-type algorithms (we remind the reader that at NLO 
the jet cross section is the same for all $k_t$-type algorithms).
The exact NLO calculation was performed here with the {\it FastJet} 
code~\cite{Cacciari:2011ma,greg}, which is also based on~\cite{Nagy}.
Kinematics are similar as before, except that we have used here the 
rapidity range $|\eta_J|\leq 0.6$. We have again used $R=0.4$ for the 
jet parameter in the $k_t$-type algorithms. 
Again, the NJA reproduces the full NLO calculation very accurately.  
Interestingly, comparing the upper and lower left parts of Fig.~\ref{fig1},
we observe that the NJA very slightly overpredicts the NLO cross section
for the case of cone algorithms, but underpredicts it in the $k_t$-type 
case. We note that the excellent overall agreement between the 
NJA and the exact NLO calculation occurs also for other choices of 
the factorization and renormalization scales. 

As an aside, we also show in the right part of Fig.~\ref{fig1}
results for similar comparisons for $p\bar{p}$ collisions at the Tevatron 
($\sqrt{S}=1960$~GeV) and for the LHC ($\sqrt{S}=7$~TeV), 
for various jet algorithms.
The kinematics correspond to those used in 
experiments~\cite{Abulencia:2007ez,Aaltonen:2008eq,CMS:2011ab,Aad:2011fc}.
The exact NLO results were again obtained using the {\it fastNLO}~\cite{fastnlo} 
and {\it FastJet}~\cite{Cacciari:2011ma,greg} packages. One can see that the 
NJA also works very well in these cases.

%%%%%%%%%%%%%%%%%%%%%%%%%%%%%%%%%%%%%%%%%%%%%%%%
\begin{figure}[t!]
\centering
\vspace*{5mm}
\includegraphics[width=8cm]{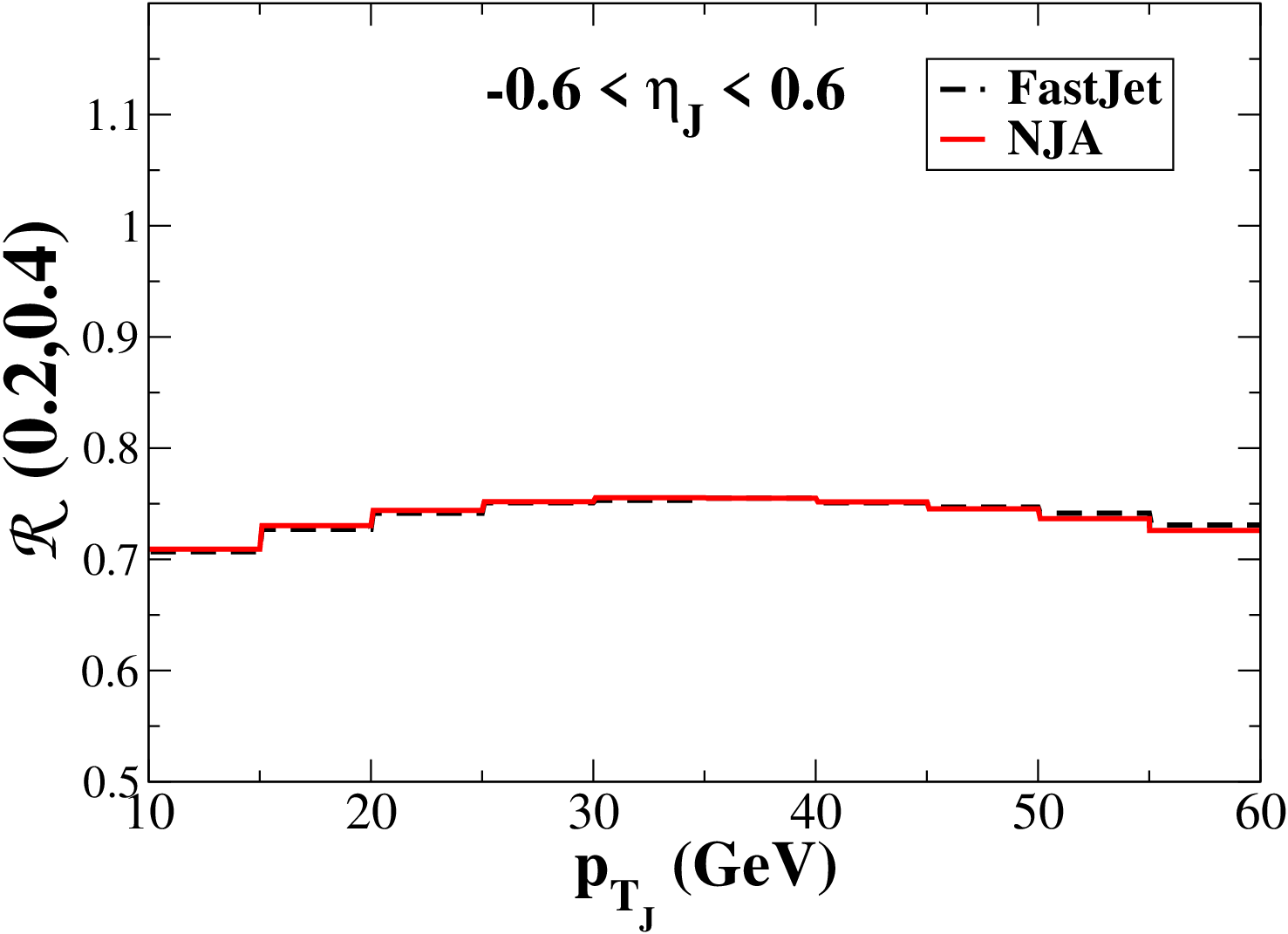}
\caption{\label{fig:ratio1}
\sf The ratio ${\cal R} (0.2, 0.4)$ as defined in Eq.~(\ref{r2}) for $pp$ collisions
at RHIC at $\sqrt{S} = 200$ GeV. The solid histogram shows our result within the NJA,
while the dashed one shows the corresponding result for the $k_t$/anti-$k_t$ algorithms presented in~\cite{Soyez:2011np}.}
\end{figure}  
%%%%%%%%%%%%%%%%%%%%%%%%%%%%%%%%%%%%%%%%%%%%%%%%
Another way of gauging the accuracy of the NJA is to consider the ratio
of jet cross sections for different jet parameters $R$:
\ba\label{r1}
{\cal R} (R_1, R_2)\equiv{{d^2\sigma/d p_{T_J}d\eta_J} (R=R_1)\over 
{d^2\sigma/ d p_{T_J}d\eta_J} (R=R_2)}.
\ea
As shown in~\cite{Soyez:2011np}, ${\cal R}$ can be expanded perturbatively 
in orders of $\alpha_s$. To the lowest non-trivial order one has
\ba\label{r2}
{\cal R}(R_1, R_2)  = 1+ \frac{d^2\sigma^{{\mathrm{NLO}}}(R_1)-
d^2\sigma^{{\mathrm{NLO}}}(R_2)}{d^2\sigma^{{\mathrm{NLO}}}|_{{\cal O}(\alpha_s^2)}},
\ea
where $d^2\sigma^{{\mathrm{NLO}}}(R)$ denotes the NLO cross section for a given $R$
and $d^2\sigma^{{\mathrm{NLO}}}|_{{\cal O}(\alpha_s^2)}$ its truncation to the
lowest order, keeping however the strong coupling constant $\alpha_s$ and the 
parton distributions at NLO. $d^2\sigma^{{\mathrm{NLO}}}|_{{\cal O}(\alpha_s^2)}$
does not depend on $R$. The difference of cross sections in the numerator of~(\ref{r2})
is of order $\alpha_s^3$, so that ${\cal R}(R_1,R_2)$ is of the form $1+{\cal O}(\alpha_s)$.
Figure~\ref{fig:ratio1} shows the result for 
${\cal R} (0.2, 0.4)$ at RHIC energy $\sqrt{S}=200$~GeV, as a function
of $p_{T_J}$. The cross sections have been integrated over $|\eta_J|\leq 0.6$,
and we have used here scales $\mu_F=\mu_R=p_{T_J}$. Our result may be directly compared
to the corresponding one given in~\cite{Soyez:2011np} for the same set of parameters,
also shown in the figure, where the full {\it FastJet} code was used.
One can see that the agreement is excellent, impressively
demonstrating the accuracy of the NJA. We note, however, that in the NJA 
the ratio ${\cal R}(R_1, R_2)$ is independent of the jet algorithm chosen.
The agreement seen in Fig.~\ref{fig:ratio1} hence is a test of the NJA as such,
but not of the implementation of a specific jet algorithm.

\begin{figure*}[t!]
\centering
\includegraphics[width=8.5cm,clip]{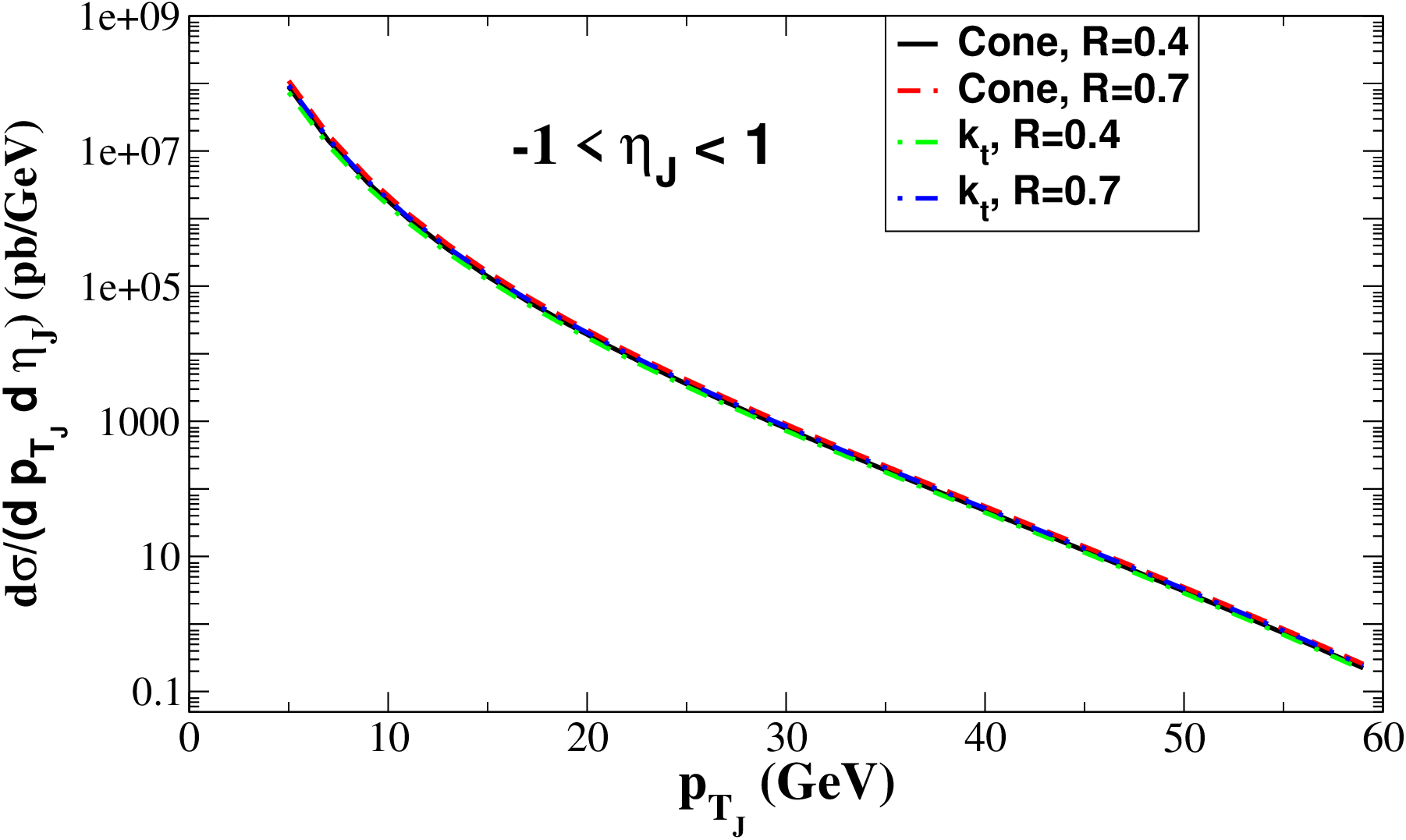}
\hspace{0.2cm}
\includegraphics[width=8.5cm,clip]{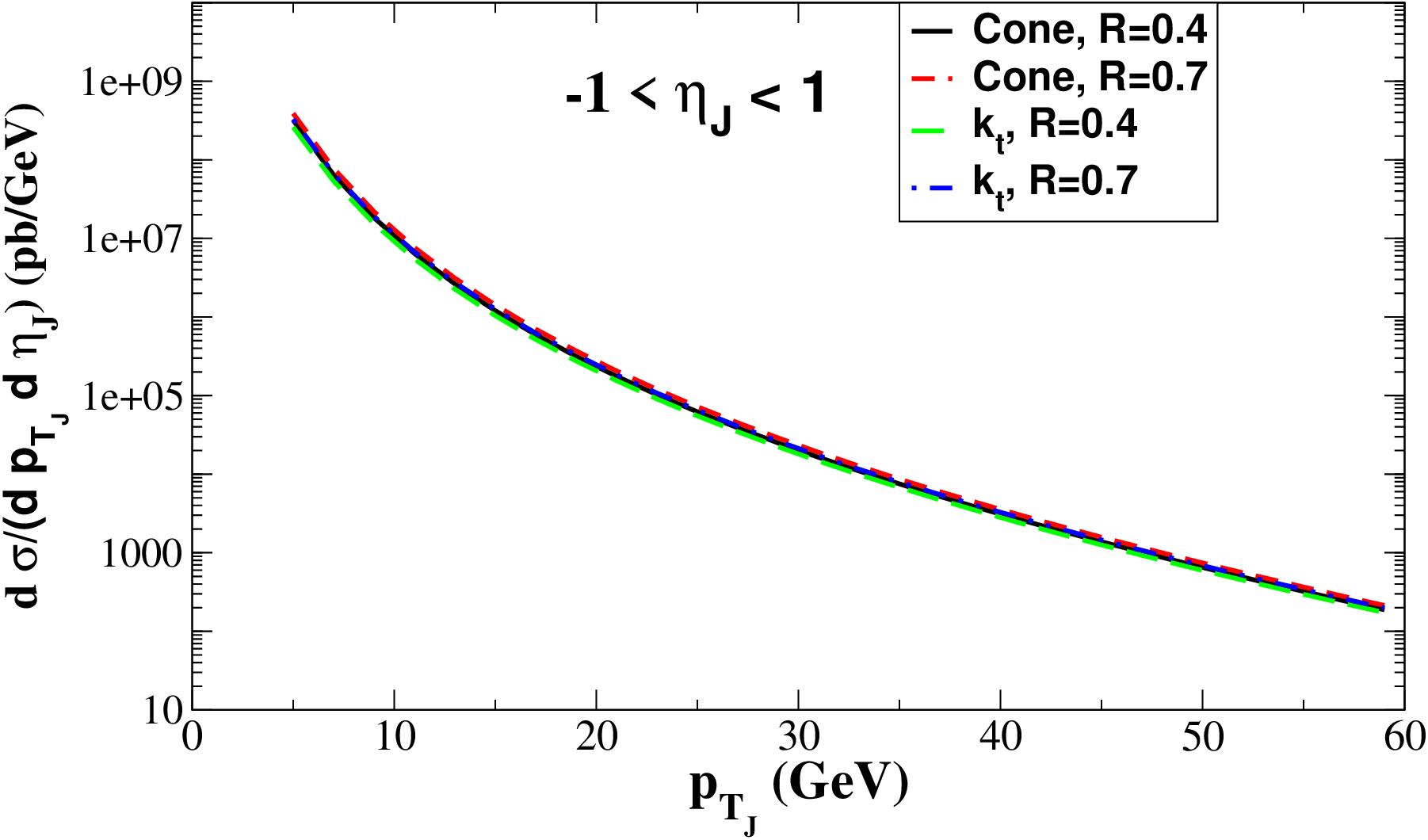}
\caption{\label{fig:cross}
\sf Spin-averaged  
NLO cross sections for single-inclusive jet production at RHIC at 
center-of-mass energies $200$~GeV (left) and~$500$~GeV (right).
Results are shown for the cone and $k_t$-type algorithms, for two 
different values of the jet parameter $R$.}
\end{figure*}

\begin{figure*}[t!]
\centering
\includegraphics[width=8.5cm,clip]{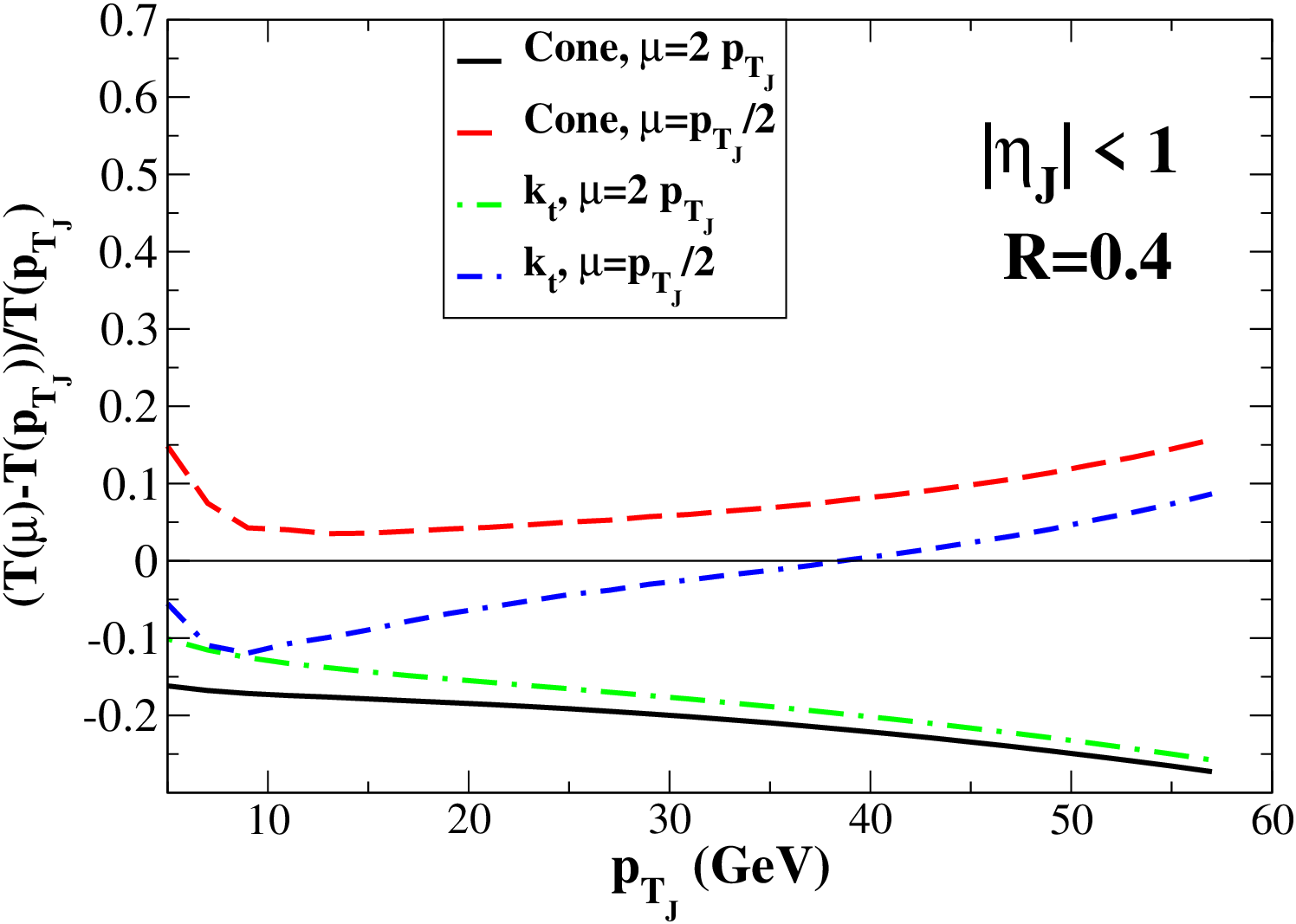}
\hspace{0.2cm}
\includegraphics[width=8.5cm,clip]{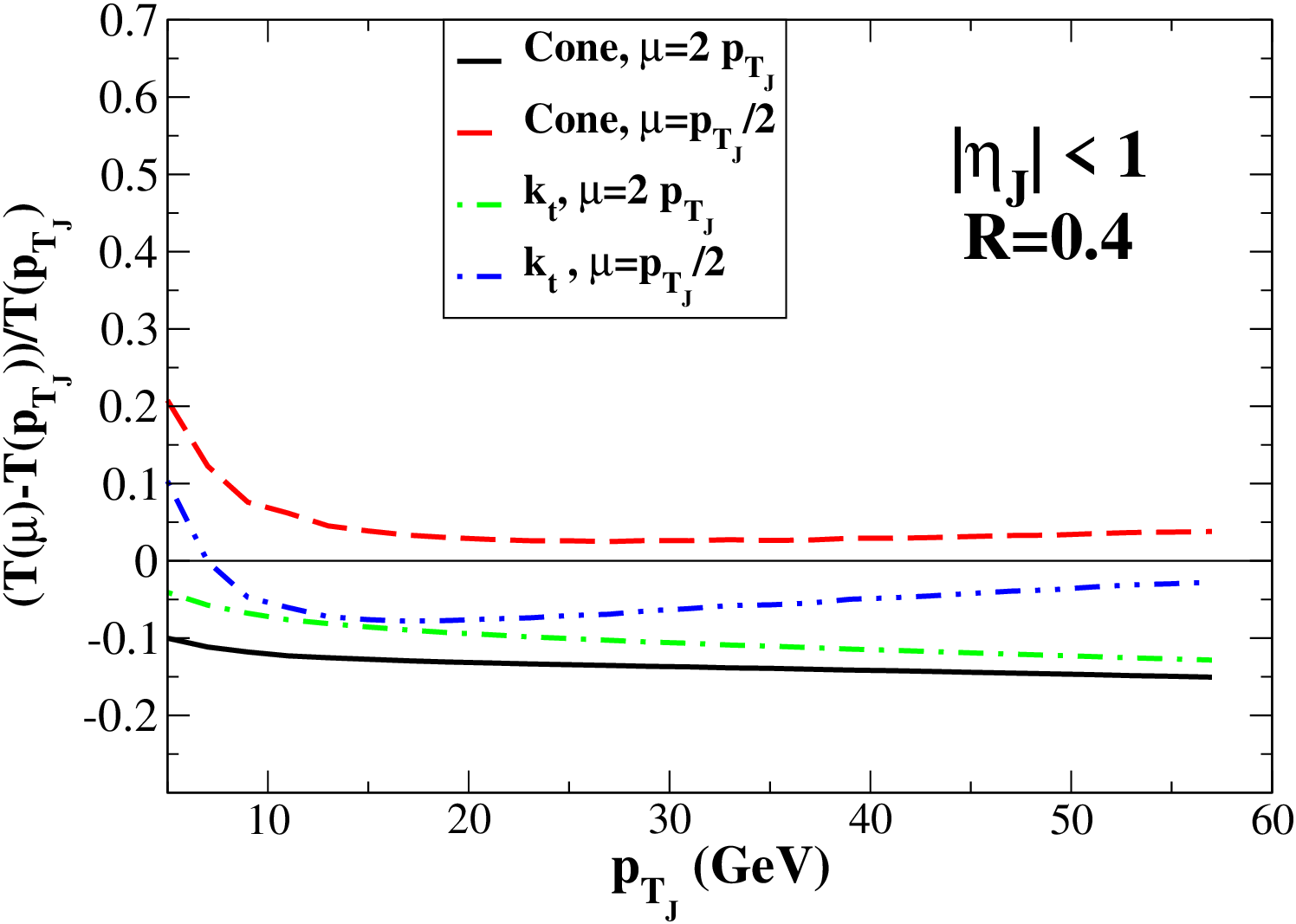}
\caption{\label{fig:scale}
\sf Scale dependence of the NLO cross sections shown in Fig.~\ref{fig:cross} 
for $R=0.4$ for center-of-mass energies $200$~GeV (left) and~$500$~GeV (right).
For notational convenience we have defined $T(\mu)\equiv d^2\sigma/dp_{T_J}d\eta_J$
at a given scale $\mu=\mu_F=\mu_R$.}
\end{figure*}

Having established the validity of the NJA, we now provide results for jet
cross sections at RHIC. Figure~\ref{fig:cross} shows the spin-averaged cross sections 
for $|\eta_J|\leq 1$ at $\sqrt{S}=200$~GeV (left) and $\sqrt{S}=500$~GeV (right). 
Results are presented for both the cone and the $k_t$-type algorithms, using
two values for the jet parameter, $R=0.4$ and $R=0.7$. The renormalization 
and factorization scales have again been set to $p_{T_J}$. Figure~\ref{fig:scale}
examines how the cross sections vary with the choice for the scale
$\mu\equiv \mu_F=\mu_R$, for the case $R=0.4$. 
We vary the scales in the region $p_{T_J}/2\leq\mu\leq 2p_{T_J}$
and show the relative deviation from the result for the case $\mu=p_{T_J}$. 
Interestingly, for this value of $R$, the scale dependence is not too similar for 
the cone and the $k_t$-type algorithms. For the former, the cross section increases 
monotonically when going from scale $\mu=2p_{T_J}$ to $\mu=p_{T_J}/2$, while
for the latter the cross section for $\mu=p_{T_J}$ is for most $p_{T_J}$ 
larger than those for both other scales. Also, the scale dependence is 
overall somewhat smaller for the $k_t$-type algorithms. We have verified 
that these patterns are present in the exact NLO
calculation with {\it FastJet}~\cite{greg}. 
%%%%%%%%%%%%%%%%%%%%%%%%%%%%%%%%%%%%%%%%%%%%%%%%
%%%%%%%%%%%%%%%%%%%%%%%%%%%%%%%%%%%%%%%%
%%%%%%%%%%%%%%%%%%%%%%%%%%%%%%%%%%%%%%%%%%%%%%%%
%%%%%%%%%%%%%%%%%%%%%%%%%%%%%%%%%%%%%%%%

\begin{figure*}[t]
\centering
\includegraphics[width=8.5cm,clip]{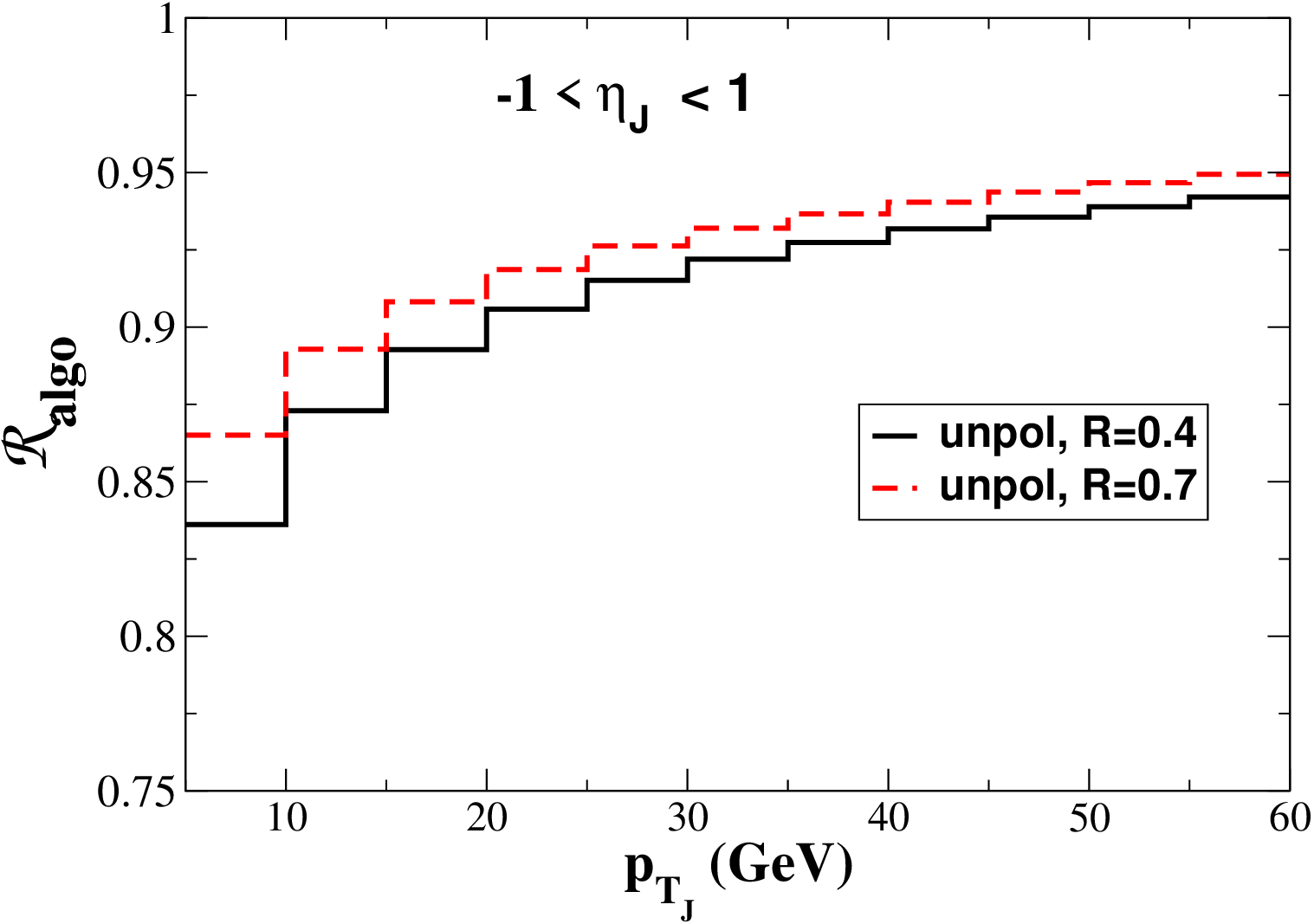}
\hspace{0.2cm}
\includegraphics[width=8.5cm,clip]{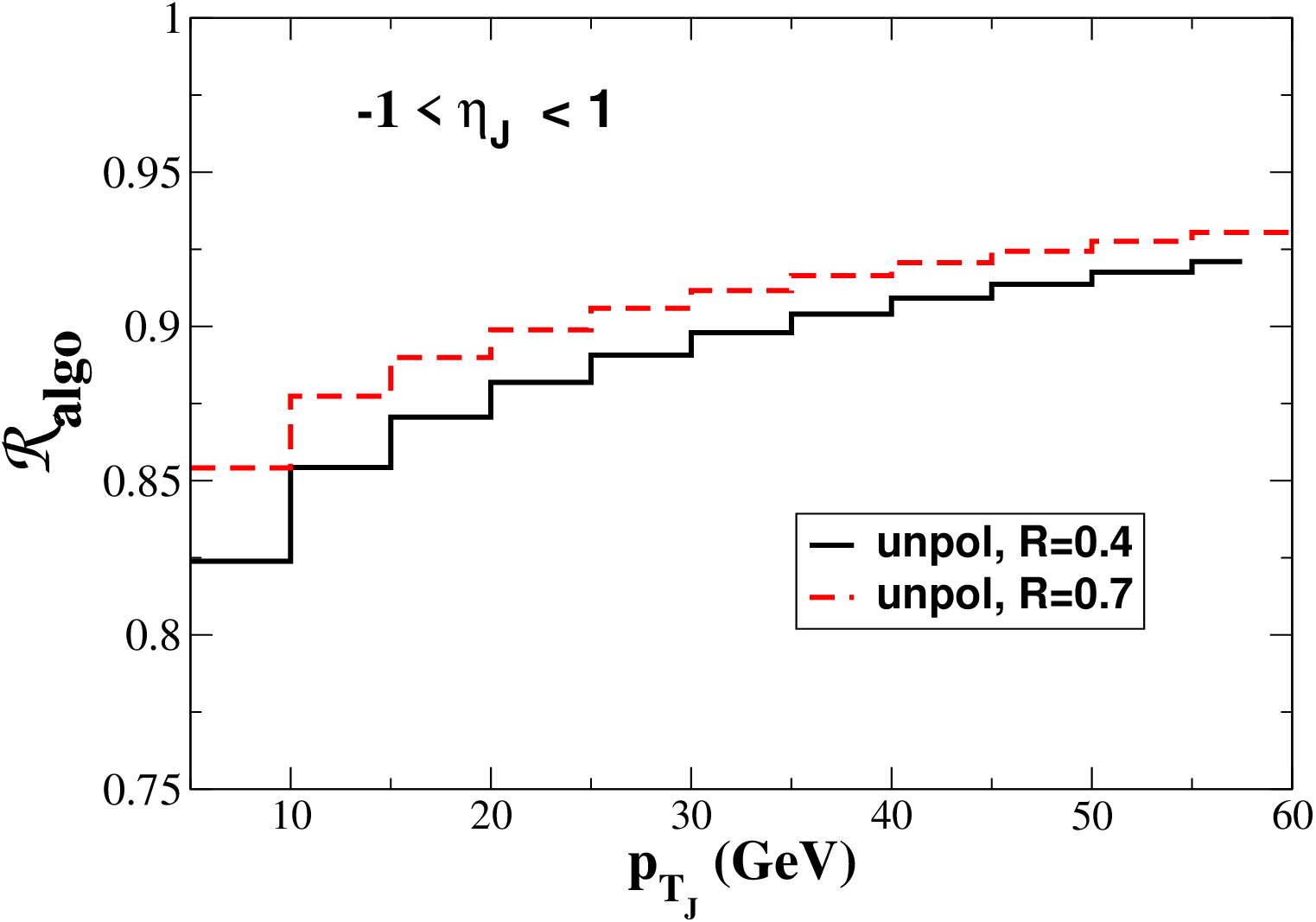}
%\vspace{-0.4cm}
\caption{\label{fig:ratio2}
\sf The ratio ${\cal R}_{\mathrm{algo}}$ at RHIC for 
$\sqrt{S} = 200$ GeV (left) and $\sqrt{S} =
500$ GeV (right), for the spin-averaged case.
Results are shown for two different values of the jet parameter $R$.}
\end{figure*}

We now turn to a more detailed comparison of the jet cross sections at RHIC 
for the two different jet algorithms. We define the ratio
\be
{\cal R}_{\mathrm{algo}} \equiv 
\frac{\left[d^2(\Delta)\sigma/d p_{T_J}d\eta_J\right]_{k_t{\mathrm{-type}}}}
{\left[d^2(\Delta)\sigma/d p_{T_J}d\eta_J\right]_{{\mathrm{cone}}}},
\ee
choosing the same jet parameter $R$ for both cross sections. Figure~\ref{fig:ratio2}
shows our results for ${\cal R}_{\mathrm{algo}}$ as a function of $p_{T_J}$
(we have chosen $p_{T_J}$ bins of 5~GeV width), for $R=0.4$ and $0.7$.
We present results for both energies relevant at RHIC, 
$\sqrt{S}=200$~GeV (left) and $\sqrt{S}=500$~GeV (right). We have in both cases
integrated the cross sections the over the pseudorapidity range $|\eta_J| \le  1$. 
We have computed the ratio for the scale $\mu=p_{T_J}$, keeping in mind, however, 
that according to Fig.~\ref{fig:scale} its scale dependence is quite large.
As one can see, the cross section for the $k_t$-type algorithms is about 10\% smaller  
than that for the cone algorithm, except at $p_{T_J}\lesssim 10$~GeV where 
the ratio ${\cal R}_{\mathrm{algo}}$ drops more strongly. Our results
are consistent with the trend seen in jet algorithm studies by STAR~\cite{star2}. 
The ratio ${\cal R}_{\mathrm{algo}}$ also shows relatively little dependence 
on the jet parameter $R$. 

The fact that the jet cross section for the $k_t$-type algorithms is found to be 
smaller than that for the cone algorithm for the same value of $R$ implies that a 
choice of a {\it larger} $R$ for the $k_t$-type algorithms should bring the 
two cross sections much closer together. Indeed, it was found 
in Ref.~\cite{catani}, that a choice $R_{k_t}\approx 1.35\, R_{\mathrm{cone}}$
makes the cross sections for the two algorithms quite similar, also for
other choices of the scale $\mu$. We confirm this finding.
%%%%%%%%%%%%%%%%%%%%%%%%%%%%%%%%%%%%%%%%%%%%%%%%
%%%%%%%%%%%%%%%%%%%%%%%%%%%%%%%%%%%%%%%%%%%%%%%%

%%%%%%%%%%%%%%%%%%%%%%%%%%%%%%%%%%%%%%%%%%%%%%%%
\begin{figure*}[t]
\vspace*{3mm}
\centering
\includegraphics[width=8.5cm,clip]{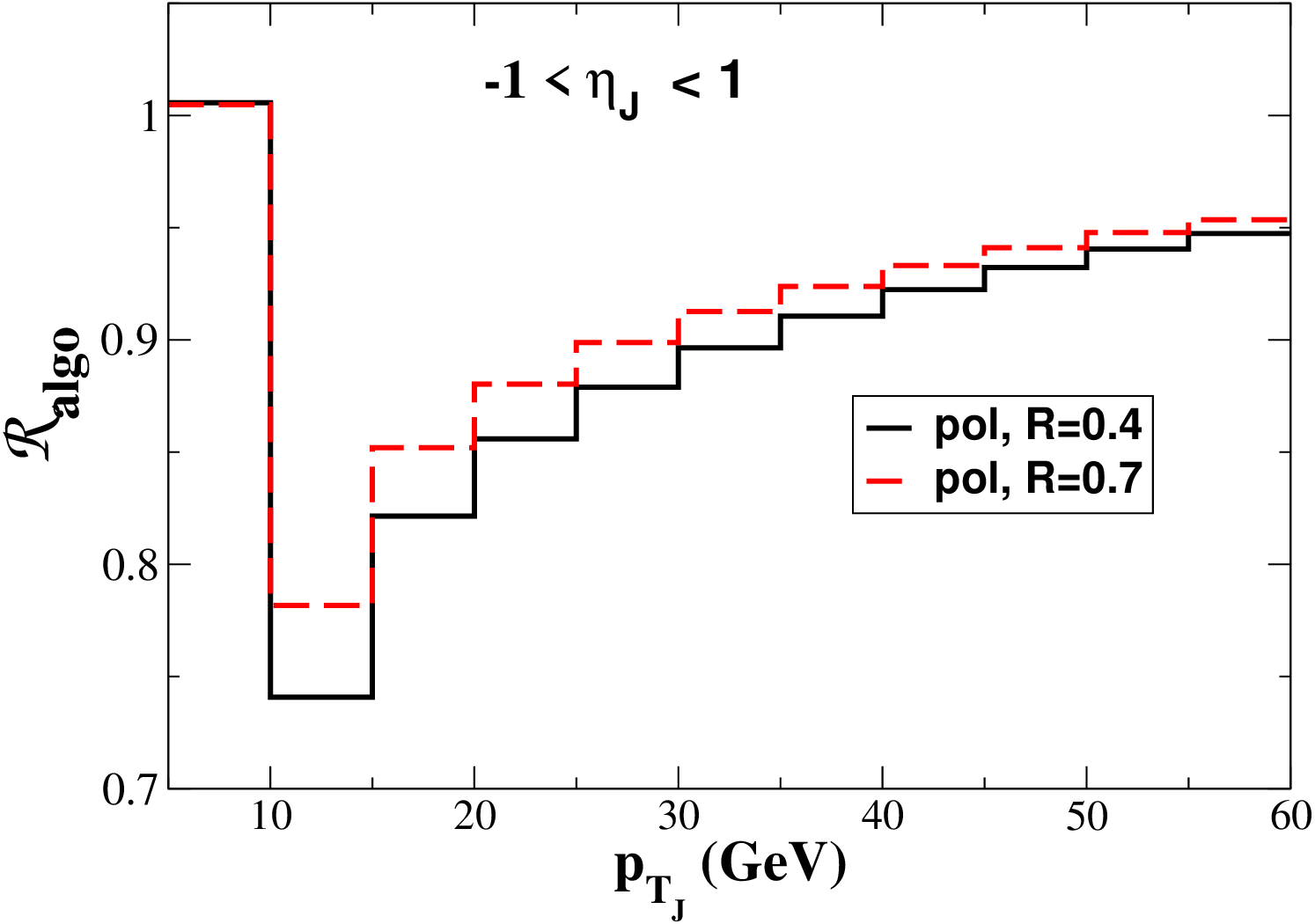}
\hspace{0.2cm}
\includegraphics[width=8.5cm,clip]{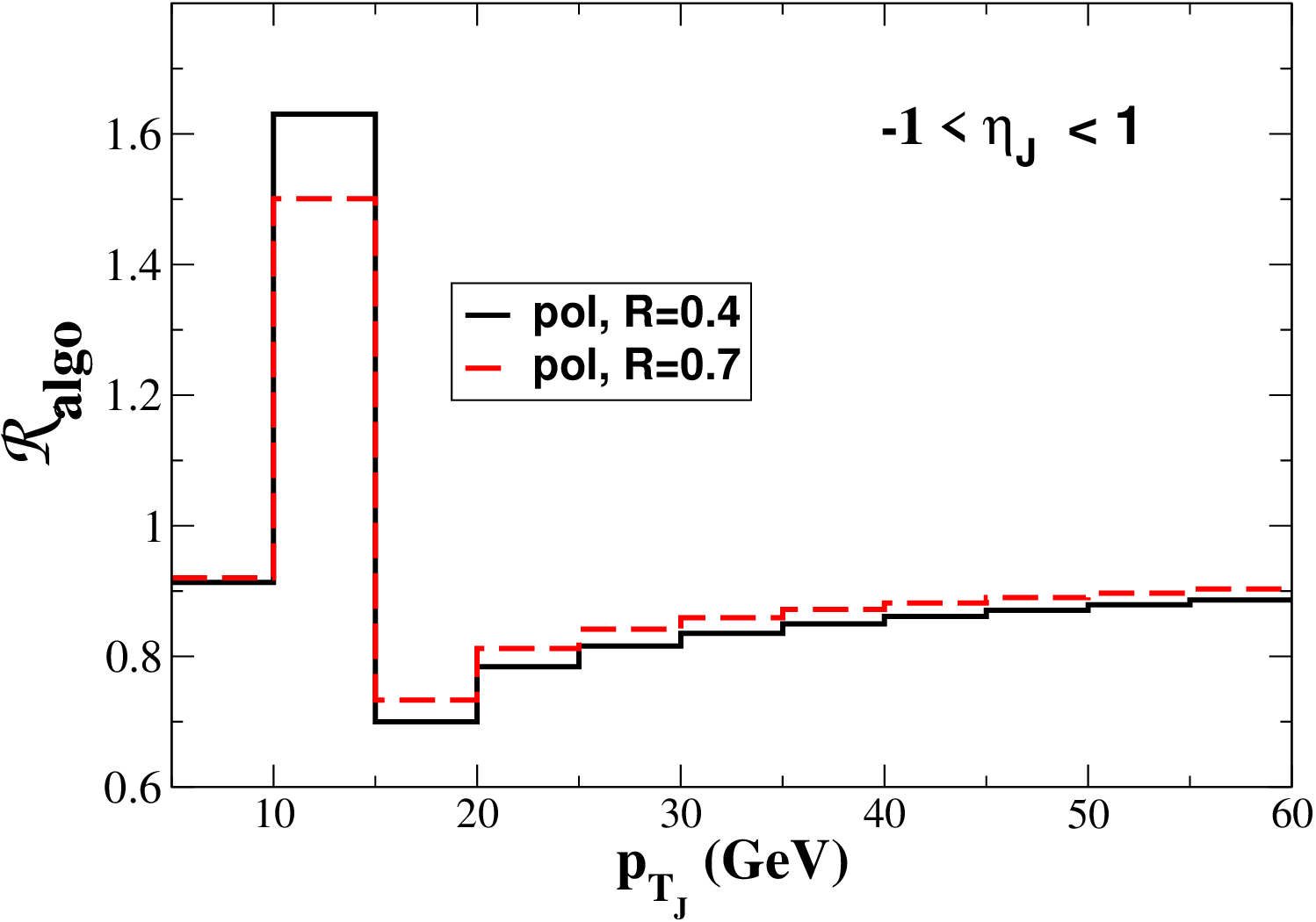}
%\vspace{-0.4cm}
\caption{\label{fig:ratio2pol}
\sf Same as Fig.~\ref{fig:ratio2}, but for the spin-dependent case.}
\end{figure*}  
%%%%%%%%%%%%%%%%%%%%%%%%%%%%%%%%%%%%%%%%%%%%%%%%

\subsection{Longitudinally polarized collisions}

For the polarized case we use the ``DSSV'' helicity parton distributions 
of Ref.~\cite{dssv}. Our first finding is that for the polarized case 
the effects of changing the jet algorithm are somewhat more pronounced
than in the unpolarized one. Figure~\ref{fig:ratio2pol} shows
the ratio ${\cal R}_{\mathrm{algo}}$ for polarized collisions at RHIC,
again computed for the scale $\mu=p_{T_J}$. ${\cal R}_{\mathrm{algo}}$
is again around 90\% at high  $p_{T_J}$, 
but shows large deviations from unity in the bin around  $p_{T_J}=12.5$~GeV. 
The reason for this is that for the DSSV set of parton distributions the
polarized jet cross section crosses zero around $p_{T_J}=10$~GeV.
Depending on the jet algorithm, this zero will be at slightly different
locations, making the denominator and numerator of ${\cal R}_{\mathrm{algo}}$
vastly different there. This is, of course, for the most part an artifact of
the way we are performing the comparison of the jet cross sections, taking
ratios of small numbers at some $p_{T_J}\sim 10$~GeV. On the other hand,
it does demonstrate the issue that in regions where the polarized cross section is
very small it may also be quite susceptible to the choice of jet algorithm
and hence (at the non-perturbative level) to hadronization corrections.
We note that the difference between the cross sections for the $k_t$-type 
and cone algorithms may again be diminished by choosing a larger value 
of $R$ for the former, $R_{k_t}\approx 1.35\, R_{\mathrm{cone}}$ as in the 
spin-averaged case discussed above. This also brings the two polarized 
cross sections somewhat closer together in the bins near their zero, even 
though marked differences remain here.

\begin{figure}[t!]
\centering
\includegraphics[width=8cm]{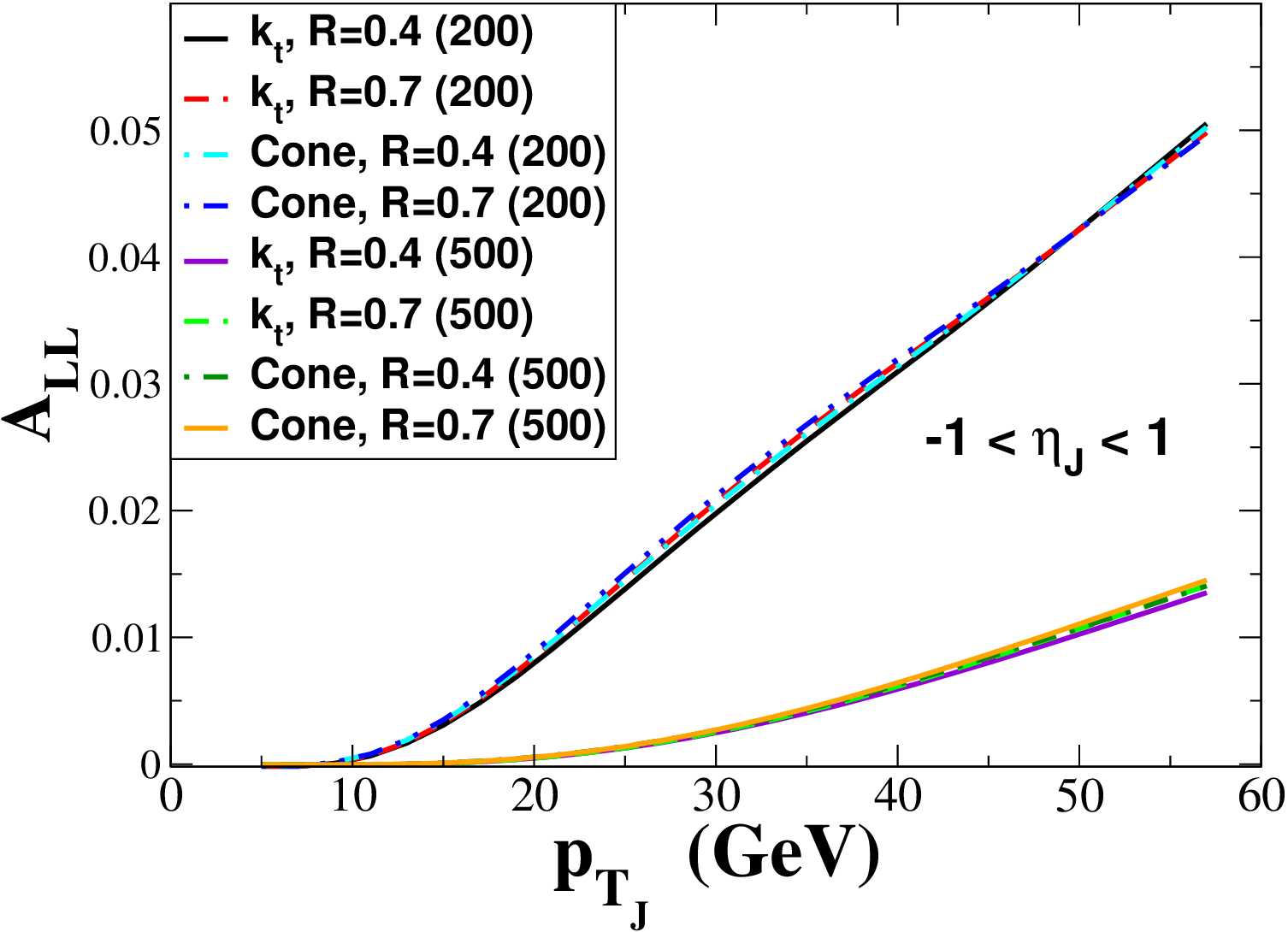}
%\vspace{-0.4cm}
\caption{\label{fig:asym}
\sf Double-longitudinal spin asymmetries $A_{LL}$ at RHIC, for $\sqrt{S}=200$~GeV and
$\sqrt{S}=500$~GeV and various jet definitions. We have averaged over $|\eta_J|\leq 1$.
The scales have been chosen as $\mu_F=\mu_R=p_{T_J}$.}
\end{figure}

Figure~\ref{fig:asym} shows the spin asymmetries $A_{LL}$ at RHIC, which are
defined by
\be\label{alldef}
A_{LL}\equiv\frac{d^2\Delta\sigma/dp_{T_J}d\eta_J}{d^2\sigma/dp_{T_J}d\eta_J}.
\ee 
For the denominator we use the spin-averaged cross sections shown in 
Fig.~\ref{fig:cross}. The most important observation is that the asymmetries
are quite insensitive to the jet algorithm chosen, and also to the value of
the jet parameter $R$. The exceptions are of course regions where the polarized
cross section (nearly) vanishes, as we saw in Fig.~\ref{fig:ratio2pol}. In these
regions, $A_{LL}$ is very small, and so these exceptions are not really noticable 
in Fig.~\ref{fig:asym}.

%%%%%%%%%%%%%%%%%%%%%%%%%%%%%%%%%%%%%%%%%%%%%
\section{Conclusions \label{sum}}
%%%%%%%%%%%%%%%%%%%%%%%%%%%%%%%%%%%%%%%%%%%%%

We have computed the NLO cross sections for single-inclusive high-$p_T$ jet production
in spin-averaged and longitudinally polarized $pp$ collisions at RHIC, with 
special focus on the algorithm adopted to define the jets. Following Ref.~\cite{jet2},
we have treated the jets in the approximation that they are rather narrow
(``Narrow Jet Approximation''). In this approximation one can derive analytical 
results for the corresponding partonic cross sections, which are of the form
 ${\cal A}\log R+{\cal B}$ with $R$ the jet parameter. 
We have extended the results of Ref.~\cite{jet2}
to the case where an infrared-safe ``$k_t$-type'' algorithm ($k_t$, anti-$k_t$, 
Cambridge/Aachen algorithm) is used.
By comparison to available ``exact'' NLO jet codes for spin-averaged 
scattering~\cite{fastnlo,Cacciari:2011ma}, we have found that the 
Narrow Jet Approximation is very accurate at RHIC for practically all 
relevant situations. The same is true even at Tevatron and LHC energies.

Our numerical results show that, for given $R$, jet cross sections at 
RHIC depend significantly on the algorithm chosen. Moreover, the scale
dependence of the cross sections can be quite different for cone- and 
$k_t$-type algorithms. For polarized cross sections, the dependence
on the jet algorithm can be very pronounced in the vicinity of a zero 
of the cross section. On the other hand, spin asymmetries at RHIC overall
turn out to be quite robust with respect to the jet algorithm adopted. 

We finally stress that our analytical results are also relevant for
matching threshold-resummed calculations of jet cross sections to 
fixed-order ones. For the case of cone algorithms, based on the results
of~\cite{jet2}, this was already exploited in~\cite{ddfwv}. Our present 
calculation allows to extend this procedure to the case of the nowadays 
more popular $k_t$-type algorithms. We note that the jets we consider here
remain massive near partonic threshold (see the discussion in~\cite{ddfwv}), which
affects the logarithmic structure of the partonic cross sections~\cite{Kidonakis:1998bk} 
and corresponds to the situation encountered in experiment and in 
the ``exact'' NLO codes such as {\it FastNLO} and {\it FastJet}.
It is known that ``non-global'' logarithms arise in this
case~\cite{Banfi:2008qs}.
%%%%%%%%%%%%%%%%%%%%%%%%%%%%%%%%%%%%%%%%%%%%%%%%
\vspace{0.cm}
%%%%%%%%%%%%%%%%%%%%%%%%%%%%%%%%%%%%%%%%%%%%%

%%%%%%%%%%%%%%%%%%%%%%%%%%%%%%%
\section{Acknowledgments}
%%%%%%%%%%%%%%%%%%%%%%%%%%%%%%%
      
We are grateful to G.~Soyez for providing 
results from his {\it FastJet} code and for helpful comments.
We also thank R.~Fatemi and C.~Gagliardi for stimulating our interest in the 
topic discussed in this paper.
AM thanks the Alexander von Humboldt Foundation, Germany, for support
through a Fellowship for Experienced Researchers. WV is grateful to 
Brookhaven National Laboratory for its hospitality during the completion 
of this work.

%%%%%%%%%%%%%%%%%%%%%%%%%%%

\end{document}